# Understanding creep of a single-crystalline Co-Al-W-Ta superalloy by studying the deformation mechanism, segregation tendency and stacking fault energy

N. Volz[a,*], F. Xue[a], C.H. Zenk[a], A. Bezold[a], S. Gabel[a], A.P.A. Subramanyam[b], R. Drautz[b], T. Hammerschmidt[b], S. K. Makineni[c,d], B. Gault[c], M. Göken[a], S. Neumeier[a]

[a] Friedrich-Alexander-Universität Erlangen-Nürnberg, Department of Materials Science & Engineering, Institute I: General Materials Properties, Erlangen, Germany

[b] Ruhr-Universität Bochum (RUB), Interdisciplinary Centre for Advanced Materials Simulation (ICAMS), Bochum, Germany

[c] Max-Planck-Institut für Eisenforschung (MPIE), Department of Microstructure Physics and Alloy Design, Düsseldorf, Germany

[d] Department of Materials Engineering, Indian Institute of Science, Bangalore, India

[*]corresponding author: N. Volz, nicklas.volz@fau.de

_______________________________________________________________________

### Abstract

A systematic study of the compression creep properties of a single-crystalline Co-base superalloy (Co-9Al-7.5W-2Ta) was conducted at 950 °C, 975 °C and 1000 °C to reveal the influence of temperature and the resulting diffusion velocity of solutes like Al, W and Ta on the deformation mechanisms. Two creep rate minima are observed at all temperatures indicating that the deformation mechanisms causing these minima are quite similar. Atom-probe tomography analysis reveals elemental segregation to stacking faults, which had formed in the γ′ phase during creep. Density-functional-theory calculations indicate segregation of W and Ta to the stacking fault and an associated considerable reduction of the stacking fault energy. Since solutes diffuse faster at a higher temperature, segregation can take place more quickly. This results in a significantly faster softening of the alloy, since cutting of the γ′ precipitate phase by partial dislocations is facilitated through segregation already during the early stages of creep. This is confirmed by transmission electron microscopy analysis. Therefore, not only the smaller precipitate fraction at higher temperatures is responsible for the worse creep properties, but also faster diffusion-assisted shearing of the γ′ phase by partial dislocations. The understanding of these mechanisms will help in future alloy development by offering new design criteria.

---

## 1. Introduction

The class of γ′ strengthened Co-based superalloys was discovered in 2006 [1] and has been studied intensively since then. These alloys are considered as alternative materials to conventionally used Ni-base superalloys for high-temperature applications [2]. Co-base superalloys are strengthened by $L1_2$-ordered precipitates as in Ni-base superalloys, however, their stoichiometry is $Co_3(Al,W)$ instead of $Ni_3Al$.

Creep properties are particularly crucial for high-temperature applications and therefore it is necessary to study the deformation mechanisms during creep. It has been shown that the deformation mechanisms of Co-base superalloys vary from the ones of CoNi- and Ni-based superalloys [3–10]. The deformation mechanisms are mainly influenced by the partitioning and segregation of alloying elements, planar fault energies and the γ/γ′ lattice misfit. The misfit is typically negative for $2^{nd} - 4^{th}$ generation single-crystal Ni-base [11–14] and positive for Co-base superalloys [15,16]. Due to their positive misfit, single-crystalline Co-base superalloys form a plate-like rafted microstructure perpendicular to the stress axis under compressive load (N-type rafting) [10,15–17], which contrasts with single-crystalline Ni-base superalloys that exhibit meander-like rafts parallel to the external compressive stress axis (P-type rafting). The behavior is vice versa if tensile stresses are considered. These microstructural observations are only valid for <001>-oriented single-crystalline material. In the following, only the case for <001>-oriented samples is considered when creep behavior or related properties are discussed. The plate-like rafted $\gamma/\gamma'$ microstructure perpendicular to the stress axis is known to strengthen both, Co-base (compressive creep) and Ni-base (tensile creep) superalloys, during creep in the high temperature creep regime [18–22]. The vertical $\gamma$ channels close because of the directional coarsening of $\gamma'$ and thus the dislocation movement becomes confined to the horizontal channels as long as the main deformation is located in the $\gamma$ matrix. This hardening effect gets lost when $\gamma'$ cutting becomes more prevalent and coarsening reduces the Orowan stress to bypass the precipitates. Xue *et al.* [18,22] found that this hardening effect of the $\gamma'$ rafts is operative in a Co-9Al-7.5W-2Ta alloy in the high temperature creep regime (950 °C / 150 MPa) and causes a second minimum in the creep rate/strain plot. This behavior is also known from Ni-base superalloys [20,23] crept in tension in the high-temperature/low-stress regime, in which they also form N-type $\gamma'$ rafts.

The main difference between Ni- and Co-base superalloys regarding the double minimum creep behavior at high temperatures is that $\gamma'$ precipitates are cut by a/2<110> dislocation pairs in Ni-base alloys [20,23,24] while the precipitate shearing in Co- and CoNi-base alloys is accompanied by the formation of extended superlattice stacking faults (SFs) [7,9,10,16–18,25–28]. Stacking faults are formed due to a thermally activated reaction of a/2<110>{111} dislocations, which results in a leading partial dislocation of a/3<112> type creating a superlattice intrinsic stacking fault (SISF) and a trailing partial dislocation of a/6<112> type that remains at the $\gamma/\gamma'$ interface. Additionally, it is known that segregation of solutes to planar faults influences the defect energies and their formation process significantly [26,29–32].

Furthermore, local diffusive reordering processes occur and defect arrangements can be changed driven by an energy reduction [30,33–35]. Another mechanism to form SISFs by climb of Frank partial dislocations was recently reported by Lenz *et al.* [26].

The change of the SISF energy due to local chemical variations has also been analyzed with density-functional theory (DFT) calculations earlier [34,36,37]. In these works, the SISF energy was approximated with the axial next-nearest neighbor Ising (ANNNI) model [38] as the energy difference of $L1_2$ and $D0_{19}$ based simulation cells with disordered Al/W sublattices [34,36,37]. With this approach, Titus *et al.* [34] found that an exchange of Al for Ta and W can reduce the stacking fault energy, which then facilitates a thermodynamically favorable phase transformation. Mottura *et al.* [36,37] reported that the addition of Ta to the $Co_3(Al,W)$ compound in general increases the SISF energy, leading to a higher resistance of $\gamma'$ against shearing. Additionally, it was calculated that a Ta atom prefers a W site rather than an Al or Co site. This preference observed for a disordered Al/W sublattice [34,36,37] was also found in corresponding DFT calculations with an ordered arrangement of atoms on the Al/W sublattice [39].

In general, all these studies found that Ta is beneficial for the mechanical properties of both Ni- and Co-base superalloys. Therefore, He *et al.* [40] investigated the same Co-9Al-7.5W-2Ta alloy as Xue *et al.* [18,22], however, at a slightly higher temperature of 975 °C. They also found a double minimum creep curve at these test parameters in compression creep as well as the formation of SFs when $\gamma'$ is cut, followed by twinning. Additionally, they showed that the SFs are enriched in the $\gamma'$ forming elements W and Ta and proposed that this promotes the formation of shear bands and twins.

Contrary to the studies of Xue *et al.* [18,22] and He *et al.* [40], Tanaka *et al.* [16], who also investigated quaternary Co-Al-W-Ta alloys in tensile creep tests at slightly higher temperatures of 1000 °C, did not find a double minimum creep behavior, although the creep regime is similar. This raises the question if the P-type rafting in tensile tests or the slightly higher temperature are responsible for this significant difference in deformation behavior since all of the effects mentioned above are diffusion-controlled processes. The diffusion coefficients are known to increase with increasing temperature in both Co- and Ni-based superalloys [41–44], suggesting faster rafting and formation of faults at higher temperatures.

To investigate the time dependency of microstructural changes during compression creep with respect to different diffusion rates, this study presents creep tests at three different temperatures. This allows the observation of kinetic effects like diffusion on the rafting behavior and the deformation mechanisms. Additionally, the experimental results of elemental segregation at a stacking fault was compared to density functional theory (DFT) calculations of the influence of solute atoms and of the Al/W ratio on the stacking fault energy. The calculations focus on the segregation of solutes on distinct sublattice sites in the planes adjacent to the SISF introduced into the supercell. Finally, the findings in this work enable to give design criteria for the future development of $\gamma'$ strengthened superalloys.

## 2. Experimental methods

### 2.1. Alloy and heat-treatment

The investigated alloy ERBOCo-2Ta, which was the focus of previous studies [18,22,40], has a nominal composition of Co-9Al-7.5W-2Ta (at.%). Single-crystal (SX) bars with a length of about 10 cm and a diameter of about 12 mm were cast in a lab-scale Bridgeman casting unit at a temperature of 1560 °C and a withdrawal rate of 3 mm/min. The bars were solution heat-treated at 1300 °C for 12 h and aged at 900 °C for 200 h followed by water quenching after each step. This heat treatment enables the formation of sufficiently large $\gamma'$ precipitates and a high $\gamma'$ volume fraction. A micrograph of the investigated alloy in the fully heat-treated state is shown in Figure 1. Additionally, several samples were heat-treated for 50 h between 900 °C to 1100 °C to evaluate the temperature dependence of the $\gamma'$ volume fraction.

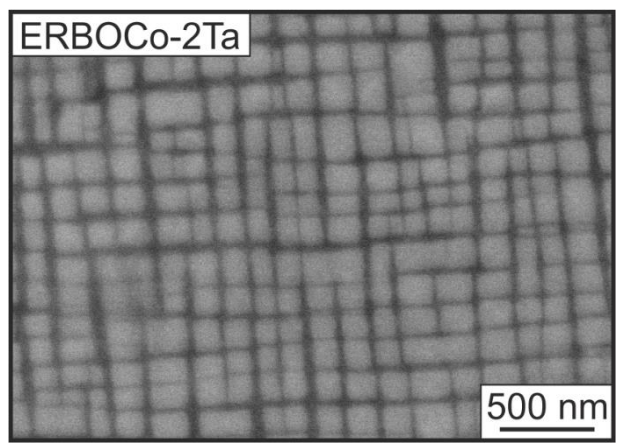

Figure 1: SEM back-scattered electron (BSE) micrograph of the investigated alloy ERBOCo-2Ta after the initial heat treatment of 1300 °C for 12 h and 900 °C for 200 h, each step followed by water quenching.

## 2.2. Microstructural characterization

Specimens for characterization by scanning electron microscopy (SEM) were stepwise ground and mechanically polished with SiC paper sheets and diamond suspensions of different particle sizes. SEM investigations were performed on a *Zeiss Cross Beam EsB 1540*. The resulting micrographs were analyzed using *ImageJ* [45] to measure γ′ area fractions and γ channel dimensions in different directions. In the style of rafting evaluations in [46], the channel dimensions were analyzed by fitting straight lines into the γ channels in binarized SEM images with a distance of 2 pixels in-between. The γ channel dimensions were determined by measuring the length of these lines. The lines were rotated in 10 ° steps to characterize the evolution of the raft formation from 0 ° to 180 ° with respect to the external load axis. The mean value of the line length at every angle was used to compare the different states. Three micrographs of different specimen areas per interrupted sample were characterized. The evaluation was performed automatically using a self-written procedure for *ImageJ*.

## 2.3. Compression creep tests

Cylindrical <001>-oriented creep specimen with a height of 7.5 mm and a diameter of 5 mm were wire eroded out of the heat-treated SX bars. The misorientation of the specimens was within 5° from the <001> direction. Compression creep tests were performed in air at a true stress of 150 MPa and temperatures of 950 °C, 975 °C and 1000 °C. The tests were repeated and interrupted at different strains and times. Two different rods were used throughout these investigations. The compression creep tests at 950 °C and 975 °C were conducted on samples from the same casting, while the tests at 1000 °C were done on a separate cast bar, leading to small differences described later in the manuscript.

Subsequently, the crept samples were cut parallel to the {001} and {010} planes to analyze the deformed microstructure. SEM samples were treated as described above. Transmission electron microscopy (TEM) samples were ground to a thickness of about 100 µm and finally electrolytically thinned using a twin-jet polishing machine with a 60 % perchloric acid in methanol and 2-butoxyethanol electrolyte (*Struers, A3 electrolyte*). TEM investigations were performed using a *Philips CM200* at 200 kV.

## 2.4. Correlative microscopy

The defect regions were located by electron channeling contrast imaging (ECCI) in a Zeiss Merlin scanning electron microscope (Carl Zeiss SMT, AG, Germany) equipped with a Gemini-type field emission gun electron column. The operating acceleration voltage was kept at 30 kV with a probe current of 4 nA. Images were taken from a working distance of 6 mm. Specimens for correlative investigation by TEM and atom probe tomography (APT) were fabricated using a dual-beam SEM / focused-ion-beam (FIB) instrument (FEI Helios Nanolab 600) via an in-situ lift-out protocol [47]. The defect regions were extracted from the bulk and subsequently attached to the electropolished tips of a halved TEM Mo-grid that was mounted in a special correlative holder designed in-house [48]. These regions were sharpened by FIB milling at 30 kV followed by a final cleaning procedure at 2 kV and 16 pA current to remove severely damaged surface areas induced previously by a high-energy Ga ion beam.

TEM on the APT needle-shaped specimens was carried out using a Philips CM-20 instrument operated at 200 kV. Near-atomic-scale compositional analysis was done by APT using a LEAP™ 5000X HR (Cameca Instruments) equipped with a reflectron lens. The instrument was operated in laser pulsing mode at a pulse repetition rate of 125 kHz and pulse energy of 40 pJ. The specimen's base temperature was kept at 60 K and the target detection rate was set to be five ions detected every 1000 pulses. Data analysis was performed using the software package IVAS 3.8.0.

To characterize and compare the segregation at the stacking fault, the concentration profiles of the alloying elements acquired by APT were fitted using a Gaussian equation. The concentrations of the element X in the $\gamma'$ phase near the stacking fault $c_{\gamma'}(X)$ and at the stacking fault $c_{SISF}(X)$ were determined from the baseline and the maximum/minimum value of the Gaussian fit, respectively. Subsequently, the defect segregation coefficient $k_x^{SISF}$ could be calculated by

$$k_X^{SISF} = \frac{c_{SISF}(X)}{c_{\gamma'}(X)}$$

(Equation 1)

## 2.5. Density functional theory calculations

The density functional theory (DFT) calculations were performed using the VASP software package [49–51] with the generalized gradient approximation to the exchange-correlation functional [52] and the projector augmented-wave method [53]. Spin-polarized calculations with a plane-wave cut off of 400 eV and a Monkhorst-Pack k-point mesh [54] with a k-point distance of 0.125 Å$^{-1}$ were performed. All calculations start from an initial ferromagnetic ordering. The electronic convergence is carried out until the total energy per ionic step converges to $10^{-6}$ eV. The atomic positions and the supercells are fully relaxed until the forces on each atom are <0.01 eV/Å. The supercells are based on $L1_2$ crystal structures with $Co_3(Al_{0.5},W_{0.5})$ (at.%) composition. In our simulations, the Al and W atoms are distributed alternatingly on the common sublattice. A complete consideration of disorder [55] would require the sampling of an intractable number of possible configurations, therefore only few arrangements of disordered Al/W sublattices are considered in this work.

The simulation supercell with a superlattice intrinsic stacking fault (SISF) is constructed from a repetition of the $L1_2$ unit cell by tilting the lattice vector in [111] direction by the Burgers vector a/3[11-2] while fixing the atomic positions. For details of the procedure of constructing the supercell, see Refs. [56–58]. The resulting SF is separated by twelve layers from its periodic images. Each layer contains eight atoms with alternating layers of Al and W atoms on the Al/W sublattice, as shown in Figure 2. In the absence of impurities, each layer has 4 Co atoms and either 2 Al atoms or 2 W atoms. This results in 12.5 % Al and 12.5 % W in the 2 layers that constitute the SISF. The segregation sites are labeled in red, on the Al sites as 1 and 4 and at the W sites as 2 and 3. The alloying element Ta is added by replacing one of the atoms of these sites. The resulting chemical composition of about 1 at.% Ta (one Ta atom in 96 atoms of the supercell) is close to the chemical composition of the alloy in the experimental part of this work.

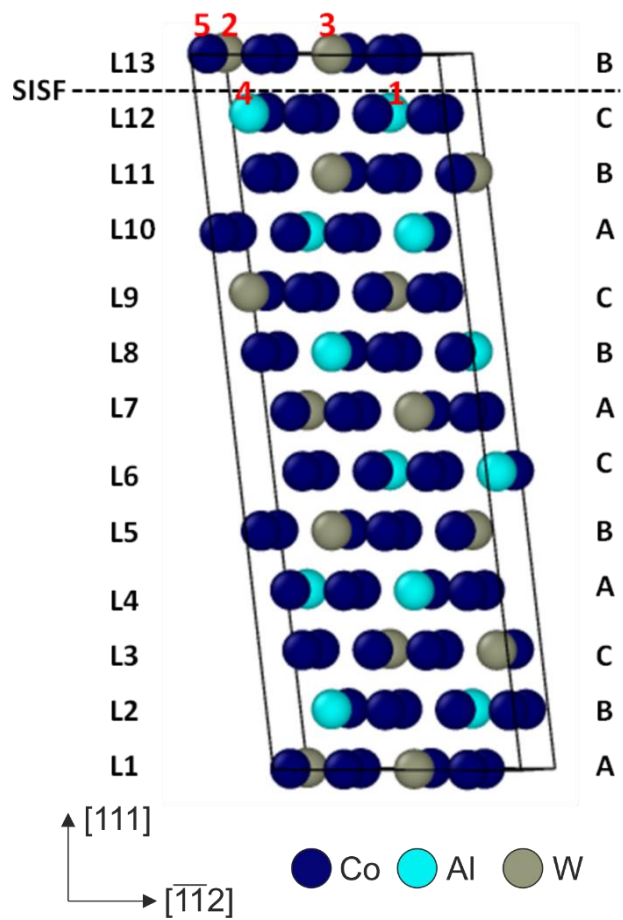

Figure 2: Supercell for DFT calculations with a SISF between layers L12 and L13 as indicated by the dashed line. The colors indicate the Co (dark blue), Al (bright blue) and W atoms (gray). The considered segregation configurations are denoted as $X_s^{(i)}$ to indicate that element $X$=Al,W,Ta replaces the atom at position $i$=1,2,3,4,5 of sublattice $s$=Co,Al,W. The layers L1 and L13 are not equivalent due to the tilt of the lattice vector.

The SISF energy is computed from the relaxed structures as

$$\gamma_{SISF} = \frac{(E_{Tot}^{SISF} - E_{Tot}^{Bulk})}{A}$$

(Equation 2)

where $E_{Tot}^{SISF}$ is the total energy of the supercell with/without alloying additions, $E_{Tot}^{Bulk}$ is the total energy of the bulk supercell in the same configuration without SF and A is the area of the SISF plane in the supercell.

## 3. Results

### 3.1. Creep properties at different temperatures

The results of the creep tests at 150 MPa and temperatures of 950 °C, 975 °C and 1000 °C are plotted in Figure 3. At 950 °C, the creep curve of ERBOCo-2Ta exhibits a very pronounced double minimum in the creep rate, showing a local minimum at strains of about 0.2 % plastic strain, followed by a slight increase of the creep rate and a global minimum at about 0.75 % strain. According to Xue *et al.* [18,22], this double minimum behavior can be attributed to the directional coarsening of γ′ during the creep test. Since the γ′ phase is an effective obstacle for the movement of a/2<110>{111} matrix dislocations at this temperature and stress, the closing of the vertical γ channels due to the formation of γ′ rafts perpendicular to the applied load results in a significant hardening [18,22].

At 975 °C, the overall creep rate significantly increases and the double minimum behavior becomes less distinct, but the two minima still appear at similar strain values. However, Figure 3b reveals the shift of the double minimum to shorter times, indicating a faster change in the mechanisms or the microstructure.

A further increase in the test temperature to 1000 °C results in higher overall creep rates. Furthermore, the hardening effect due to the formation of γ′ rafts is less visible since especially the first minimum in the strain rate plots becomes less pronounced. However, other microstructural features like the marginally larger precipitate size due to the higher temperatures of 1000 °C might also have an influence on the diminishing creep rate minima. The minimum creep rates are reached at approximately similar strains compared to the tests at 950 °C and 975 °C. However, especially the first minimum is shifted to shorter times, as shown in Figure 3b.

The measurements indicate that a temperature shift from 950 °C to 975 °C leads to a more pronounced influence on the creep curves compared to the shift from 975 °C to 1000 °C.

However, as described above, samples from two separate castings were used. The samples used for the tests at 1000 °C exhibit a slightly higher $\gamma'$ fraction at equal temperatures, assumingly caused by small variations in the overall chemical composition. The effect of the $\gamma'$ volume fraction is discussed in detail in section 4, since a higher $\gamma'$ fraction leads to a more pronounced strengthening, which results in lower creep rates [59]. However, the double minimum creep and the underlying mechanisms are still clearly visible at all temperatures and in all samples.

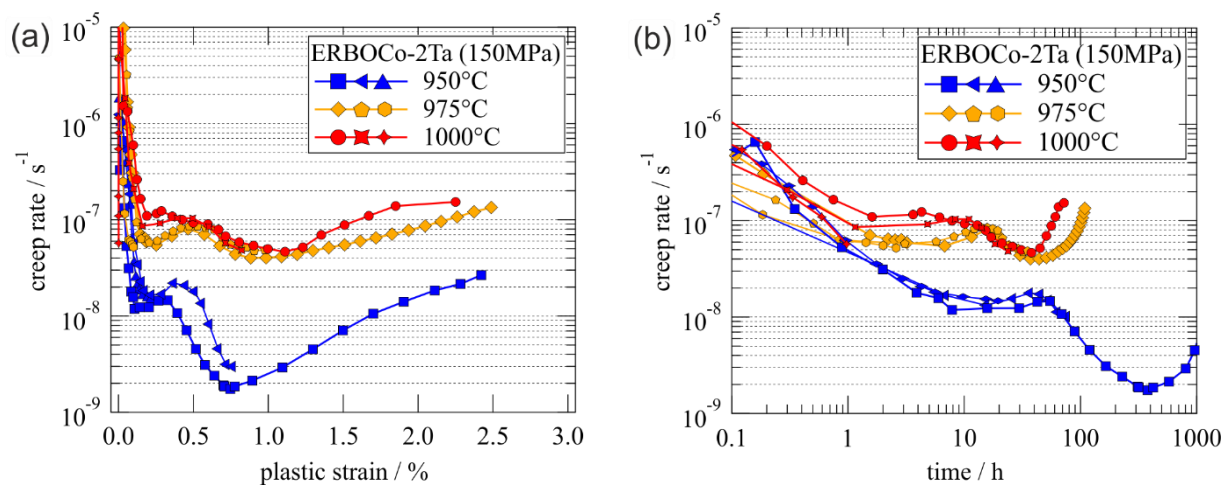

Figure 3: Creep tests of ERBOCo-2Ta at 950 °C, 975 °C and 1000 °C / 150 MPa. (a) Creep rate plotted against plastic strain and (b) creep rate plotted against the creep time.

## 3.2. Microstructural evolution during creep

The creep tests were repeated and interrupted after several strains/times (see Figure 3) to investigate the evolution of the $\gamma/\gamma'$ microstructure. Figure 4 shows the corresponding SEM micrographs.

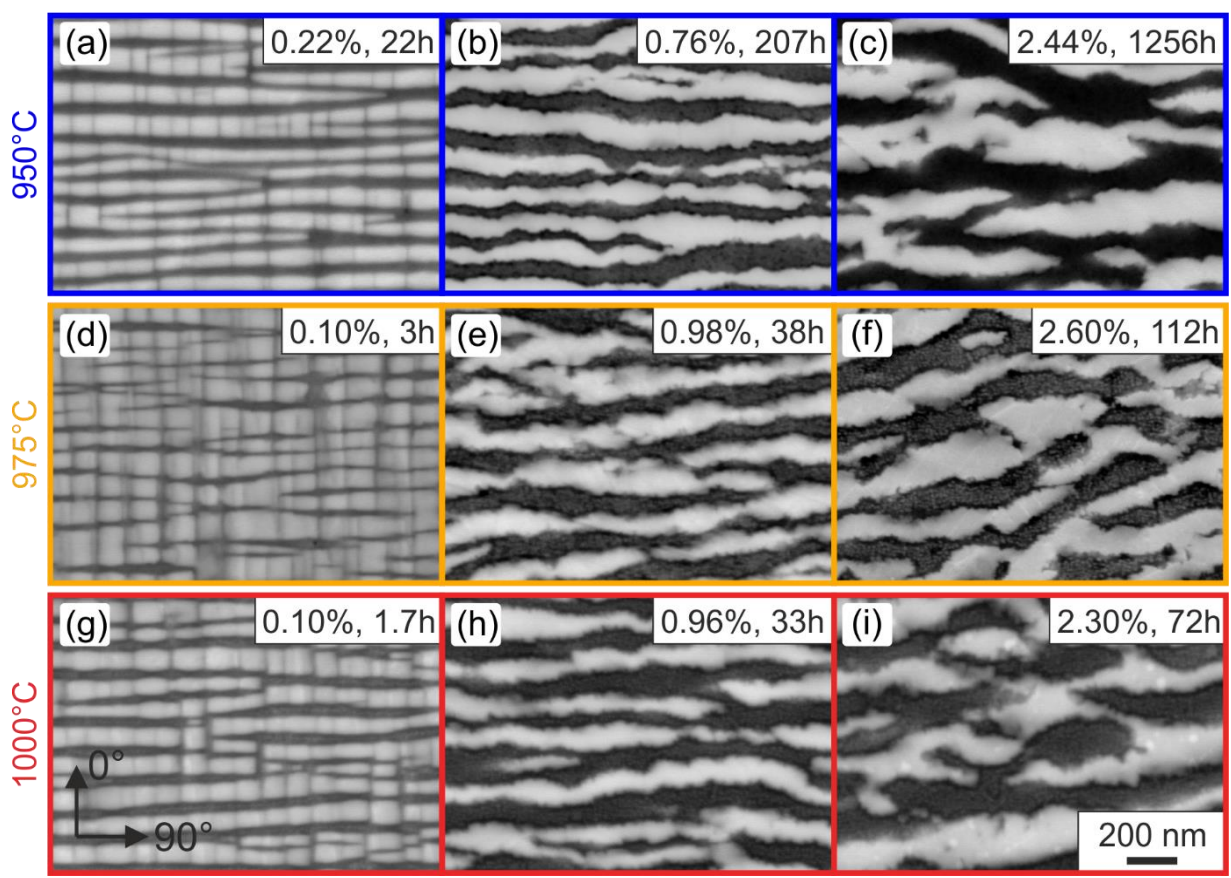

Figure 4: SEM microstructure after creep tests at 150 MPa and (a)-(c) 950 °C, (d)-(f) 975 °C and (g)-(i) 1000 °C. Tests were interrupted after different plastic strains/times.

A pronounced directional coarsening of γ' is visible at all test temperatures, caused by the high γ/γ' lattice misfit in combination with external loading. This rafting behavior was quantified by measuring the γ channel dimension in several directions, as illustrated in Figure 5. Due to the positive lattice misfit and the applied compressive stress, the preferred rafting direction is perpendicular to the external stress axis. At later stages of the creep tests, the γ' phase becomes irregularly shaped and grows in other directions as well. This re-arrangement or rotation of the rafts is well described by He *et al.* [40]. They propose a mass transport of γ' forming solutes by partial dislocations shearing through γ' on <111> planes, which leads to the re-precipitation of γ' in the γ matrix on <111> planes when the dislocations are stopped at γ/γ' interfaces. Since rafting is also a diffusion-controlled process, an increasing temperature results in a faster evolution of the rafted structure. The specimen tested at 975 °C to a strain of about 0.98 % (Figure 4 (e) and Figure 5 (b)), for example, exhibits γ channel widths comparable to the sample tested at 950 °C to a similar strain of about 0.76 % (Figure 4 (b) and Figure 5 (a)), however, the test time was only 38 h compared to over 200 h. This is even more pronounced when one compares the samples tested until 2.44 % (Figure 4 (c)) and 2.3 % (Figure 4 (i)) strain at 950 °C and 1000 °C, respectively. Both samples show comparable γ channel dimension, while the test time differs in a factor of more than 17. In summary, the evolution of the directional coarsening is similar at comparable strain values, independent of the test temperature, as indicated by Figure 5. This significant faster formation of the rafted structure at higher temperatures is mainly attributed to the higher diffusivity of the alloying elements.

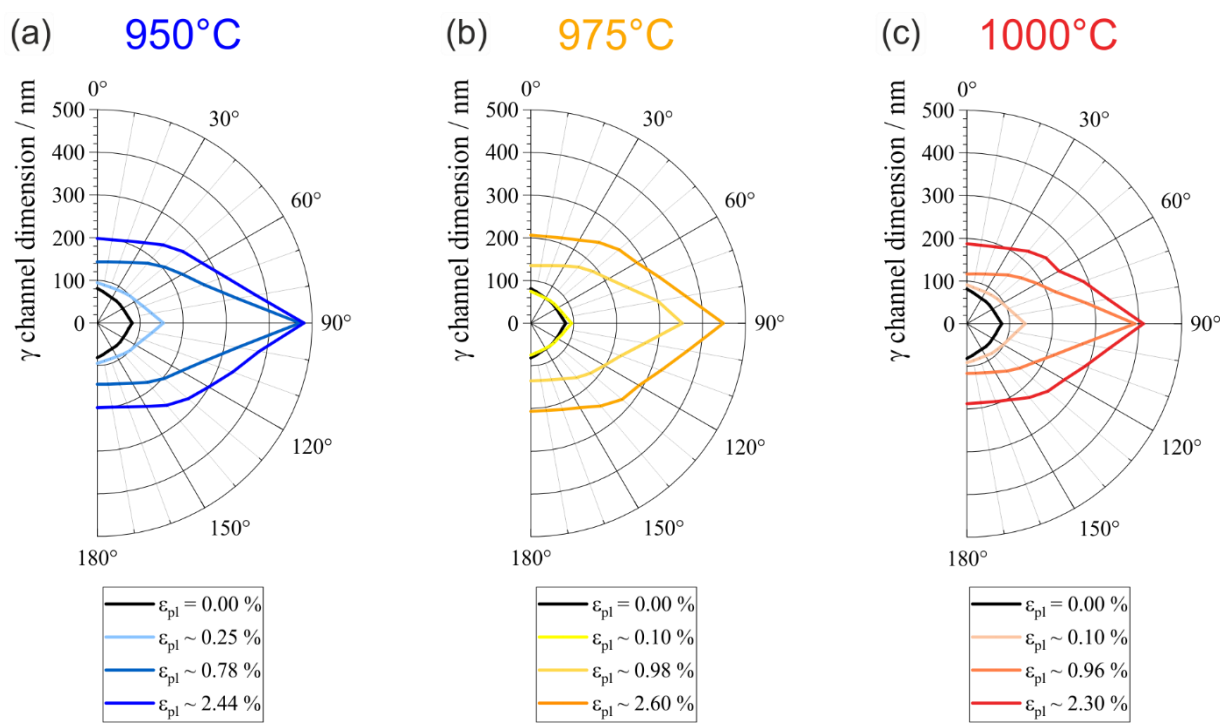

Figure 5: Quantification of the γ channel dimensions during creep at the temperatures (a) 950 °C, (b) 975 °C and (c) 1000 °C. The undeformed microstructure is given in black as 0.0 % plastic strain. The angles refer to the coordinate system introduced in Figure 4.

Additionally, the samples were investigated by TEM and ECCI after the interrupted creep tests. The corresponding defect structures are shown in Figure 6.

It is evident that the deformation at 950 °C up to strains of 0.22 % and 0.76 % is mainly located in the γ channels since long dislocation segments are visible in the horizontal γ channels (see Figure 6 (a)) and stacking faults within γ′ were observed rarely. After about 0.76 % strain, the γ′ precipitates are sheared more frequently. However, deformation is still concentrated in the γ matrix, as indicated by the high dislocation density and the dense interfacial dislocation network shown in Figure 6 (b) and discussed in detail elsewhere [22]. At larger strains of 2.44 %, significant shearing of γ′ and the formation of stacking faults can be observed (see Figure 6 (c)). This indicates that the hardening effect of γ′ vanishes as the creep test proceeds, leading to a pronounced softening as visible in the creep data in Figure 3. Twinning and the formation of small $D0_{19}$ nuclei could be observed at this stage. However, this will not be discussed in detail in this manuscript and the reader is referred to Xue *et al.* [22] for more information.

At 975 °C and low strain/short time of 0.1 % and 3 h (Figure 6 (d)), the microstructure looks quite similar to the same stages at 950 °C (Figure 6 (a)). However, after about 1 % strain (see Figure 6 (e)), shearing events under stacking fault formation are observed more frequently. The microstructure after 2.6 % strain at 975 °C (see Figure 6 (f)) is again comparable to 950 °C (see Figure 6 (c)) and the precipitates are frequently cut by dislocations, resulting in a high stacking-fault density inside the plate-like γ′ rafts. Additionally, twinning was observed in this crept sample. This mechanism is explained in detail by He *et al.* [40].

The deformation mechanisms at 1000 °C are comparable to the ones at 975 °C and 950 °C. Cutting is only observed infrequently at low strains and the main deformation is concentrated in the γ matrix phase (see Figure 6 (g)). At large strains of 2.3 % (Figure 6 (i)), the rafts are packed with stacking faults and cutting of γ′ is a relevant mechanism. However, at the intermediate stage of about 1 % plastic strain at 1000 °C (Figure 6 (h)), γ′ cutting is pronounced already, similar to the microstructure after 2.3 % (Figure 6 (i)). This indicates that cutting of γ′ occurs earlier when the temperature is higher.

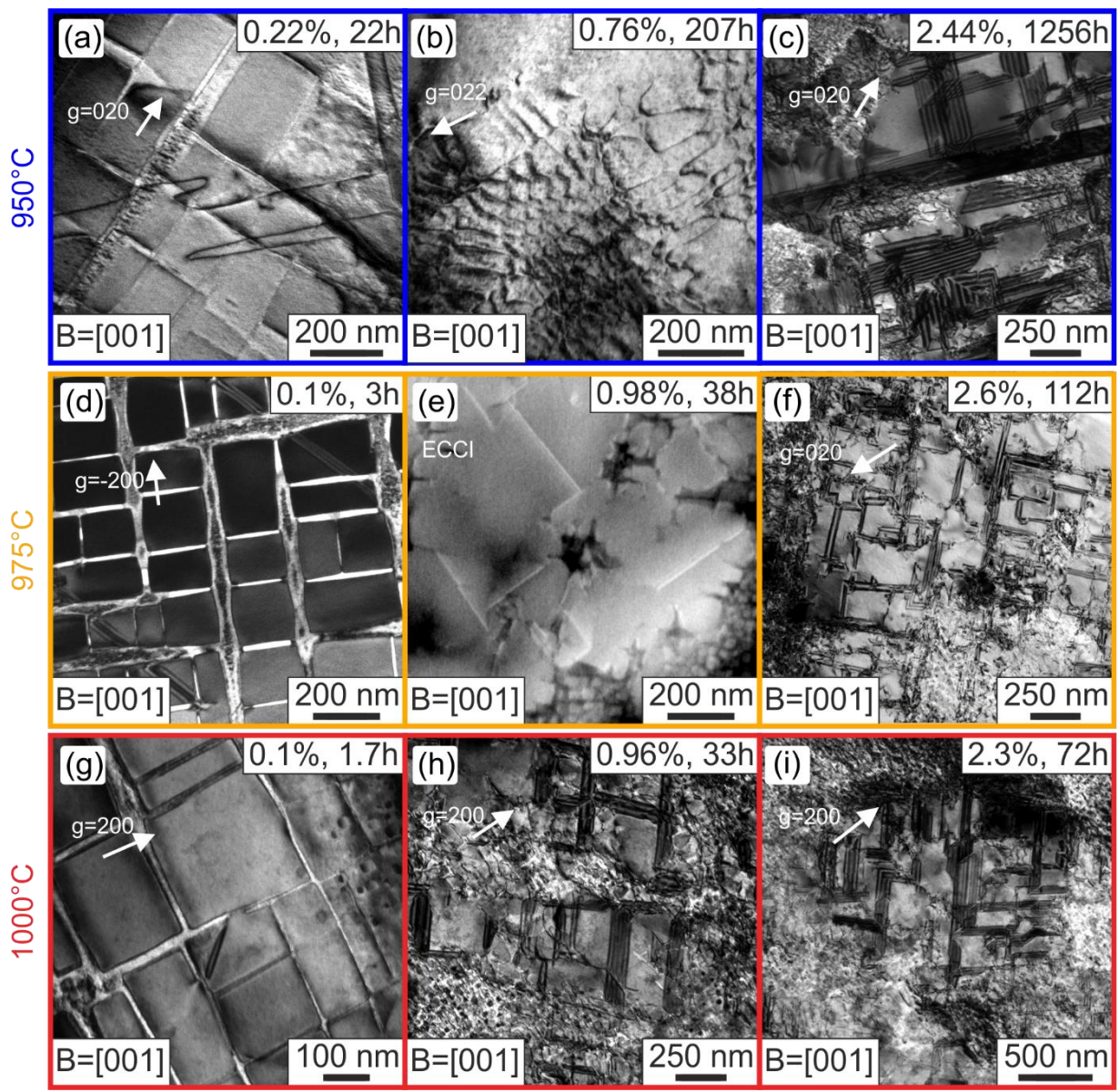

Figure 6: TEM/ECCI microstructures after creep tests at 150 MPa and (a)-(c) 950 °C, (d)-(f) 975 °C and (g)-(i) 1000 °C. Tests were interrupted after different plastic strains/times.

### 3.3. Elemental segregation at stacking faults

An APT specimen was prepared from the sample crept to about 1 % plastic strain at 975 °C, in order to investigate the segregation of solutes to creep-deformation-induced stacking faults. As evident from the TEM micrograph in Figure 7 (a), the APT specimen contained a stacking fault in the γ′ phase. The corresponding reconstructed element map (see Figure 7 (b)), showing Al atoms in red and an iso-compositional surface delineated by 55 at.% Co in yellow to mark the γ/γ′ interface, indicates that the distribution of Al changes along the stacking fault. The composition profile across the stacking fault shown in Figure 7 (c) reveals a depletion in Co, a significant depletion in Al and an enrichment of W and Ta. The quantification of the segregation behavior is given in Table 1. The defect segregation coefficient $k_W^{SISF}$ of 1.28 for W and $k_{Ta}^{SISF}$ of 1.32 for Ta indicate that their degree of segregation towards the SISF is similar. The $k_{Al}^{SISF}$ of 0.70 shows that the depletion of Al is much more pronounced compared to the depletion of Co ($k_{Co}^{SISF} = 0.98$).

Table 1: Concentration of the solutes at the SISF and in the surrounding γ′ phase determined from APT and the calculated defect segregation coefficient.

|  | Co | Al | W | Ta |
|---|---|---|---|---|
| $c_{SISF}$ / at.% | 75.47 | 6.34 | 13.43 | 4.56 |
| $c_{γ'}$ / at.% | 76.89 | 9.08 | 10.53 | 3.45 |
| $k_X^{SISF}$ / - | 0.98 | 0.70 | 1.28 | 1.32 |

Similar TEM-EDS and APT results were reported by Titus *et al.* [28,34] for an Co-8.8Al-9.8W-2.0Ta alloy crept at 900 °C and 345 MPa. Further investigations on the local compositions around SISFs in a multi-component CoNi-base superalloys (Co-32Ni-8Al-5W-6Cr-2.5Ti-1.5Ta-0.4Si-0.1Hf) also find Al depletion and Ta and W enrichment but contrary to the findings reported herein, Co was reported to be enriched around the stacking fault [26,35]. However, a direct comparison is not meaningful in this case because of the presence of additional elements and complex solute interactions. In particular the Ni-content is known to have a large influence on the stacking fault energy of γ [60] and γ′ [7].

These findings indicate that segregation of solutes to and away from faults during creep already occurs at comparably short times and low strains. Since higher temperatures increase the diffusivity of the atoms, the segregation to faults is likely to happen faster during the creep tests at 975 °C and 1000 °C compared to the tests at 950 °C. However, the fault energy itself is temperature dependent and so the driving force for segregation might also be different.

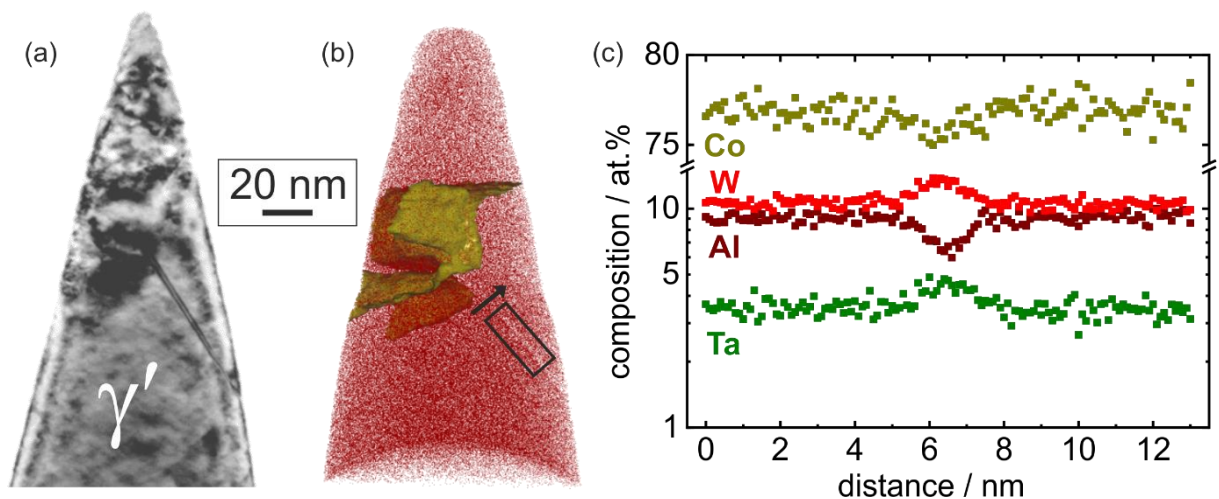

Figure 7: APT specimen in (a) TEM and (b) after reconstruction (Al atoms shown in red; Co 55 at.% iso-concentration surface in yellow, the arrow indicates the area of the concentration profile) showing (c) Co/Al depletion and W/Ta enrichment at the stacking fault. This sample was prepared from a specimen creep-deformed for 38 h at 975 °C to 0.98 % plastic strain.

### 3.4. DFT calculations of stacking fault energies

The segregation profiles of the APT experiments in Figure 7 can be directly compared to the DFT calculations of the changes in SISF energy $\gamma_{SISF}$ with respect to segregation in the vicinity of the fault plane. The corresponding supercells are generated by placing W, Al or Ta atoms at Al sites (site 1 and 4 in Figure 2), W sites (site 2 and 3 in Figure 2) or Co-sites (site 5 in Figure 2). Figure 8 (a) shows the computed SISF energies for the pure SISF supercell, i.e. without segregated atoms, and for different segregation configurations. In order to compare our explicit calculations of the SISF energy to the previous estimates with the DFT+ANNNI approach [34,36,37], we include not only the results for the fully relaxed SISF simulation cells in Figure 8 (a), but also for the unrelaxed SISF simulation cells. The simulation cells of the latter were generated by the introduction of a SF in the bulk cell as described above without further relaxation.

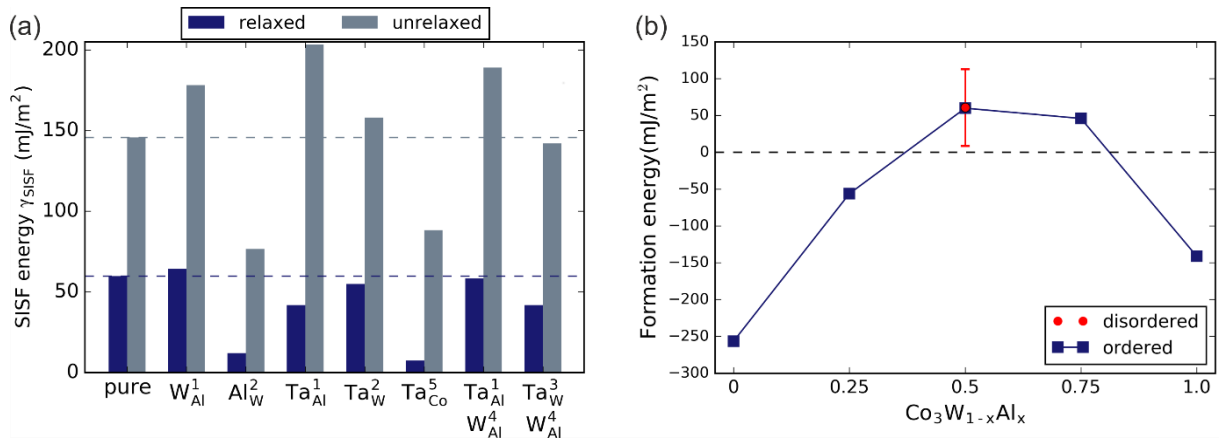

Figure 8: (a) SISF energies for different segregation configurations as obtained from DFT calculations with unrelaxed (gray) and fully relaxed (blue) supercells. The x-axis denotes the sublattice (subscript) and particular site (superscript) that the segregating atom occupies as introduced in Figure 2. (b) SISF energies of $Co_3W_{1-x}Al_x$ without Ta ('pure') for different Al/W ratios with an ordered (blue) Al/W sublattice. The value of the disordered Al/W sublattice (red) is taken as the average with standard deviation of five randomly chosen disordered arrangements.

Overall, a considerable difference owing to full relaxation of the supercell is observed with respect to both the order of magnitude and individual qualitative differences. Our value for the relaxed SF energy in the pure, ordered $Co_3(Al,W)$ of 60 mJ/m$^2$ is slightly lower than the value of 89-93 mJ/m$^2$ of the DFT+ANNNI estimate [34] that used a disordered Al/W sublattice but no explicit treatment of the stacking fault. The difference of approximately 30 mJ/m$^2$ is of the order of the mixing energy in this system [55] and considerably smaller than the effect of atomic and volume relaxation of nearly 90 mJ/m$^2$. The very good agreement with the average of five randomly chosen disordered arrangements of the Al/W sublattice (Figure 8 (b)) confirms the validity of our approach to analyze the SF energies with ordered Al/W sublattices. A much larger influence on the SF energies is observed due to atomic relaxations: The calculations without atomic relaxation show an increase of the SISF energy for W replacing Al at the fault plane ($W^1_{Al}$), for Ta replacing Al or W at the fault plane ($Ta^1_{Al}$, $Ta^2_W$) and for a combination thereof ($Ta^1_{Al}$ plus $W^4_{Al}$). The previous DFT+ANNNI estimates of the SISF energy [34,36,37] are to some degree comparable to these results in the sense that they neglect the influence of atomic relaxation due to the presence of the SF. Similar to our results with unrelaxed atomic positions, the configuration-averaged DFT+ANNNI estimate indicates an increase of the SISF energy for Ta on an Al or W site [34]. Full relaxation of the supercell with SF however, leads to qualitatively different results in that the SISF energy is decreased for Ta on an Al or W site ($Ta^1_{Al}$, $Ta^2_W$, $Ta^1_{Al}$ plus $W^4_{Al}$). As the total change in volume and shape of the supercell before and after relaxation is less than 1 %, this demonstrates that the small energy changes due to the influence of the SF on atomic relaxation can alter the qualitative picture and interpretation.

Our DFT calculations with atomic relaxation can rationalize the segregation profiles of Co and Ta measured by APT experiments (see Figure 7). In the calculations, the addition of a Ta atom next to the fault reduces the SISF energy, regardless if the Ta atom replaces a Co, Al, or W atom. The largest reduction is predicted for Ta replacing Co ($Ta^5_{Co}$), followed by Ta replacing Al ($Ta^1_{Al}$) and W ($Ta^2_W$). Interestingly, this sequence is opposite to previously calculated energetic sequence of solution energies [34,35]. The preference of Ta replacing Co at the SF plane matches the Ta enrichment and Co depletion at the SF plane observed in APT. In order to analyze the influence of the experimentally observed W enrichment and Al depletion at the SF plane, we additionally determined the SISF energy for Ta on an Al or W site with co-segregation of a W atom on an Al site. The SISF energies of the corresponding configurations $Ta^1_{Al}$ plus $W^4_{Al}$ as well as $Ta^3_W$ plus $W^4_{Al}$ are similar to Ta on Al ($Ta^1_{Al}$) or W ($Ta^2_W$) without W on Al co-segregation and higher than Ta replacing Co. We therefore expect that the preference of Ta for the Co sites at the SF plane is not affected by the local W over Al enrichment.

Our DFT calculations for $W^1_{Al}$ and $Al^2_W$ indicate an increase of the SISF energy for W replacing Al and a decrease for Al replacing W in agreement with previous DFT calculations [61]. These DFT results of replacing single atoms at the SF plane can, however, not explain the APT experiment in this work and in previous works [34] that show an enrichment of W and a depletion of Al at the SF plane. We therefore conducted additional DFT calculations to analyze the influence of the Al/W ratio, as shown in Figure 8 (b). The Al/W chemical compositions of 3:1 and 1:3 are represented with alternating arrangement of W and Al atoms every second layer in $Co_3Al$ and $Co_3W$ supercells, respectively, in the supercells with and without SF. Our results in Figure 8 (b) show that the SISF energy is reduced for deviations from a 1:1 Al/W ratio towards both, Al and W enrichment, in agreement with previous DFT calculations [61]. For $Co_3W$ and $Co_3Al$ the SISF energy takes negative values, which is not surprising as the B2 phase in Co-Al and the $D0_{19}$-phase in Co-W are competing in structural stability at these compositions. For small deviations from 1:1 composition, the SISF energy is lowered by a small amount for Al enrichment and by a considerable amount for W enrichment. This is in line with the W enrichment and the Al depletion at the SISF observed in APT (Figure 7). The observed enrichments at the SISF are hence due to different thermodynamic driving forces that arise from local atomic relaxations in the case of Ta and from a deviation of chemical stoichiometry in the case of W.

## 4. Discussion

To evaluate the kinetic differences in the evolution of the creep rates during the creep tests on ERBOCo-2Ta at 950 °C, 975 °C and 1000 °C, all the consequences of the results presented above need to be considered. First of all, an increasing temperature leads to an increasing diffusivity of the alloying elements [41–43] and a decreasing γ′ volume fraction. Especially the lower γ′ fraction of 56 % at 975 °C compared to 69 % at 950 °C is known to significantly reduce the creep resistance in Co-base [59] and Ni-base [62] superalloys and can explain the overall higher creep rates at higher temperatures. At 1000 °C, the specimens exhibit a γ′ fraction of about 58 %, which results from the different cast rod, as, described above, leading to a smaller difference between the creep properties at 975 °C and 1000 °C.

However, also the time dependency of the deformation mechanisms is changing at the different temperatures, *i.e.* the double minimum curve gets less distinct and is shifted to shorter times at 975 °C and 1000 °C, compared to 950 °C. In the literature, it is suggested that the directional coarsening of the γ′ phase perpendicular to the applied load can work as a strengthening mechanism by hampering the dislocation movement [18–23]. However, this can only occur in the high temperature/low stress regime as long as the deformation is predominantly located in the γ phase. Additionally, the N-type rafting of the γ′ phase seems to be responsible for the double minimum in the creep tests of Co- and Ni-base superalloys in the high-temperature regime under compressive and tensile load, respectively [18,22,63]. When the vertical γ channels are closing due to the raft formation, the γ′ rafts act as considerable obstacles for the dislocation movement and higher stresses and a longer pathway are required to overcome the precipitate phase, which results in a second decrease of the creep rate. Additionally, interactions of stacking faults and twins in the γ′ phase were reported in [22]. Therefore, interaction of dislocations with SFs or twins on other glide planes cannot be excluded as a hardening mechanism besides directional coarsening. Furthermore, slight differences in precipitate fractions and sizes can have an influence on the mechanical properties and mechanisms. However, we assume that the cutting of the γ′ phase by dislocations at later stages of the creep tests is responsible for the softening of the material.

The behavior described above was observed for ERBOCo-2Ta at 950 °C and 150 MPa and was also found at 975 °C and 1000 °C and 150 MPa. Interestingly, in all tests, the first and second minimum of the creep rates can be observed at similar plastic strains of about 0.2 % and 0.75 %, respectively. However, this means that the minima are reached at significantly shorter times at 975 °C and 1000 °C, since the overall creep rates are higher (see Figure 3). The rafted microstructures look nearly identical at the same strain but different times, since the diffusion velocity is higher at the higher temperatures and the raft formation is faster. That a certain strain is needed to enable directional coarsening was already reported earlier [64]. Even if the γ/γ′ lattice misfit and thus the driving force for rafting of ERBOCo-2Ta is smaller at higher temperatures (see [18]), the directional coarsening is faster. This means that the hardening effect due to closing vertical channels happens faster at higher temperatures, as can be seen in the creep data in Figure 3. Of course, the softening due to intensive cutting and the faster coarsening of $\gamma'$ also appears at shorter times. Maybe these findings also explain why Tanaka *et al.* [16] did not find a double minimum during tensile creep tests at comparable temperature and strain, and instead only observed a sharp minimum followed by a plateau in the creep rate. Of course, also the N-type raft formation of the alloys in [16] might cause the difference in the deformation behavior.

Additionally, the shift of shearing of $\gamma'$ to shorter times is promoted by the segregation of alloying elements to stacking faults in $\gamma'$. It is known that segregating elements around faults are a significant factor in the formation and propagation of defects [31,32]. Segregation to stacking faults was confirmed in the ERBOCo-2Ta sample creep-deformed to 1 % plastic strain at 975 °C (Figure 7) and was also shown earlier in a sample crept to longer times at this temperature (112 h, 2.6 %) [40]. It was quantified that the degree of segregation is identical for Ta and W, whereas the depletion of Al is more pronounced compared to Co (Figure 7). DFT calculations, as shown in Figure 8, show that the stacking fault energy is reduced by segregation of Ta and W to the fault, leading to an easier formation and extension of the present superlattice stacking faults.

It is known that segregation can not only be found at SFs but also at different  types of dislocations [26,31,35,40,65]. Since it is assumed that the deformation during the early stages of creep is located in the $\gamma$ matrix and dislocations are arranged at $\gamma/\gamma'$ interfaces, the local reduction of the stacking fault energy can also promote cutting since the necessary energy for a dislocation to cut into $\gamma'$ is then also reduced. This can be seen in Figure 6 at the micrographs after about 1 % plastic strain. While at 950 °C, cutting is not very prominent, the amount of stacking faults due to dislocations shearing through $\gamma'$ is increasing significantly at higher temperatures. This shows, that the γ′ precipitates are sheared after much shorter times due to the faster segregation kinetics. Therefore, the softening effect is shifted to shorter times at similar plastic strain as the test temperature increases.

For the future development of Co-base superalloys, several design criteria can be formulated based on this study. Since the deformation is predominantly located in the γ matrix in the early creep stages, a low stacking fault energy of the matrix phase is beneficial. A large dissociation distance of dislocations hampers their recombination and therefore cross slip. However, a high superlattice stacking fault energy is needed for the γ′ phase to prevent partial dislocations from cutting into the precipitates. If alloying elements are added that segregate to stacking faults and reduce the stacking fault energy, they should be slow diffusers. Additionally, a high positive γ/γ′ lattice misfit is beneficial during compressive creep in the high temperature creep regime, since the directional coarsening perpendicular to the stress axis is an effective strengthening mechanism.

## 5. Summary and conclusion

A systematic study of the creep properties of ERBOCo-2Ta in the high-temperature regime revealed the time-dependent deformation mechanisms during creep and their relation to diffusive effects. The following conclusions can be drawn:

- Two creep rate minima in the creep curve are visible at all temperatures, indicating that similar deformation mechanisms are active.

- Rafting is significantly faster at higher temperatures, leading to a shift of the double creep rate minimum to shorter times, while the plastic strains corresponding to the minima are roughly identical.

- The softening due to cutting of γ′ is also shifted to shorter test times at higher temperatures, since the faster segregation to dislocations and stacking faults facilitates γ′ shearing.

- DFT calculations using a fully relaxed simulation supercell with an SISF revealed different thermodynamic driving forces for the segregation of Ta and W towards the SISFs. The calculated SISF energy changes due to local segregation are in good agreement with the experimentally determined elemental distribution around a SISF.

**Acknowledgment**

The authors acknowledge funding by the Deutsche Forschungsgemeinschaft (DFG) through projects A4, B3 and C1 of the collaborative research center SFB/TR 103 "From Atoms to Turbine Blades – a Scientific Approach for Developing the Next Generation of Single Crystal Superalloys". SKM acknowledges Humboldt foundation for the fellowship. FX acknowledges funding by the Sino-German (CSC-DAAD) Postdoc Scholarship. U. Tezins, C. Broß and A. Sturm are acknowledged for their support to the FIB and APT facilities at the Max-Planck-Institut für Eisenforschung.

**Literature**

[1]    J. Sato, Cobalt-Base High-Temperature Alloys, Science. 312 (2006) 90–91. https://doi.org/10.1126/science.1121738.
[2]    J.L. Cann, A. De Luca, D.C. Dunand, D. Dye, D.B. Miracle, H.S. Oh, E.A. Olivetti, T.M. Pollock, W.J. Poole, R. Yang, C.C. Tasan, Sustainability through alloy design: Challenges and opportunities, Prog. Mater. Sci. (2020) 100722. https://doi.org/10.1016/j.pmatsci.2020.100722.
[3]    A. Suzuki, T.M. Pollock, High-temperature strength and deformation of γ/γ' two-phase Co–Al–W-base alloys, Acta Mater. 56 (2008) 1288–1297. https://doi.org/10.1016/j.actamat.2007.11.014.
[4]    D.P. Pope, S.S. Ezz, Mechanical properties of Ni3Al and nickel-base alloys with high volume fraction of γ', Int. Met. Rev. 29 (1984) 136–167. https://doi.org/10.1179/imtr.1984.29.1.136.
[5]    M. Feller-Kniepmeier, T. Link, Correlation of microstructure and creep stages in the< 100> oriented superalloy SRR 99 at 1253 K, Metall. Mater. Trans. A. 20 (1989) 1233–1238.
[6]    T.M. Pollock, A.S. Argon, Creep resistance of CMSX-3 nickel base superalloy single crystals, Acta Metall. Mater. 40 (1992) 1–30. https://doi.org/10.1016/0956-7151(92)90195-K.
[7]    M.S. Titus, Y.M. Eggeler, A. Suzuki, T.M. Pollock, Creep-induced planar defects in L12-containing Co- and CoNi-base single-crystal superalloys, Acta Mater. 82 (2015) 530–539. https://doi.org/10.1016/j.actamat.2014.08.033.
[8]    Y.M. Eggeler, M.S. Titus, A. Suzuki, T.M. Pollock, Creep deformation-induced antiphase boundaries in L12-containing single-crystal cobalt-base superalloys, Acta Mater. 77 (2014) 352–359. https://doi.org/10.1016/j.actamat.2014.04.037.
[9]    H.J. Zhou, H. Chang, Q. Feng, Transient minimum creep of a γ' strengthened Co-base single-crystal superalloy at 900 °C, Scr. Mater. 135 (2017) 84–87. https://doi.org/10.1016/j.scriptamat.2017.03.031.

[10] F. Xue, H.J. Zhou, Q.Y. Shi, X.H. Chen, H. Chang, M.L. Wang, Q. Feng, Creep behavior in a γ′ strengthened Co–Al–W–Ta–Ti single-crystal alloy at 1000 °C, Scr. Mater. 97 (2015) 37–40. https://doi.org/10.1016/j.scriptamat.2014.10.015.

[11] H. Biermann, M. Strehler, H. Mughrabi, High-temperature X-ray measurements of the lattice mismatch of creep-deformed monocrystals of the nickel-base superalloy SRR99, Scr. Metall. Mater. 32 (1995) 1405–1410.

[12] S. Neumeier, F. Pyczak, M. Göken, The temperature dependent lattice misfit of rhenium and ruthenium containing nickel-base superalloys – Experiment and modelling, (2020) submitted.

[13] T.M. Pollock, R.D. Field, Chapter 63 Dislocations and high-temperature plastic deformation of superalloy single crystals, in: Dislocations Solids, Elsevier, 2002: pp. 547–618. https://doi.org/10.1016/S1572-4859(02)80014-6.

[14] R.C. Reed, The Superalloys: Fundamentals and Applications, (n.d.) 390.

[15] F. Pyczak, A. Bauer, M. Göken, S. Neumeier, U. Lorenz, M. Oehring, N. Schell, A. Schreyer, A. Stark, F. Symanzik, Plastic deformation mechanisms in a crept L12 hardened Co-base superalloy, Mater. Sci. Eng. A. 571 (2013) 13–18. https://doi.org/10.1016/j.msea.2013.02.007.

[16] K. Tanaka, M. Ooshima, N. Tsuno, A. Sato, H. Inui, Creep deformation of single crystals of new Co–Al–W-based alloys with fcc/L1 $_2$ two-phase microstructures, Philos. Mag. 92 (2012) 4011–4027. https://doi.org/10.1080/14786435.2012.700416.

[17] M.S. Titus, A. Suzuki, T.M. Pollock, Creep and directional coarsening in single crystals of new γ–γ′ cobalt-base alloys, Scr. Mater. 66 (2012) 574–577. https://doi.org/10.1016/j.scriptamat.2012.01.008.

[18] F. Xue, C.H. Zenk, L.P. Freund, M. Hoelzel, S. Neumeier, M. Göken, Double minimum creep in the rafting regime of a single-crystal Co-base superalloy, Scr. Mater. 142 (2018) 129–132. https://doi.org/10.1016/j.scriptamat.2017.08.039.

[19] D.-W. Chung, D.S. Ng, D.C. Dunand, Influence of {$\gamma^\prime$}-raft orientation on creep resistance of monocrystalline Co-based superalloys, Materialia. 12 (2020) 100678. https://doi.org/10.1016/j.mtla.2020.100678.

[20] R.C. Reed, N. Matan, D.C. Cox, M.A. Rist, C.M.F. Rae, Creep of CMSX-4 superalloy single crystals: effects of rafting at high temperature, Acta Mater. 47 (1999) 3367–3381.

[21] U. Tetzlaff, H. Mughrabi, Enhancement of the high-temperature tensile creep strength of monocrystalline nickel-base superalloys by pre-rafting in compression, Pollock TM Kissinger RD Bowman RR Green KA McLean M. (2000). http://www.tms.org/superalloys/10.7449/2000/Superalloys_2000_273_282.pdf (accessed August 8, 2017).

[22] F. Xue, C.H. Zenk, L.P. Freund, S. Neumeier, M. Göken, Understanding raft formation and precipitate shearing during double minimum creep in a γ′-strengthened single crystalline Co-base superalloy, Philos. Mag. (2020) accepted for publication.

[23] H. Mughrabi, W. Schneider, V. Sass, C. Lang, The Effect of Raft Formation on the High-Temperature Creep Deformation Behaviour of the Monocrystalline Nickel-Base Superalloy CMSX-4, (n.d.) 4.

[24] A. Epishin, T. Link, Mechanisms of High Temperature Creep of Nickel-Base Superalloys Under Low Applied Stress, in: Superalloys 2004 Tenth Int. Symp., TMS, 2004: pp. 137–143. https://doi.org/10.7449/2004/Superalloys_2004_137_143.

[25] Y.M. Eggeler, J. Müller, M.S. Titus, A. Suzuki, T.M. Pollock, E. Spiecker, Planar defect formation in the γ′ phase during high temperature creep in single crystal CoNi-base superalloys, Acta Mater. 113 (2016) 335–349. https://doi.org/10.1016/j.actamat.2016.03.077.

[26] M. Lenz, M. Wu, E. Spiecker, Segregation-assisted climb of Frank partial dislocations: An alternative route to superintrinsic stacking faults in L12-hardened superalloys, Acta Mater. 191 (2020) 270–279. https://doi.org/10.1016/j.actamat.2020.03.056.

[27] N. Volz, C.H. Zenk, R. Cherukuri, T. Kalfhaus, M. Weiser, S.K. Makineni, C. Betzing, M. Lenz, B. Gault, S.G. Fries, J. Schreuer, R. Vaßen, S. Virtanen, D. Raabe, E. Spiecker, S. Neumeier, M. Göken, Thermophysical and mechanical properties of advanced single crystalline Co-base superalloys, Metall. Mater. Trans. A. 49 (2018) 4099–4109. https://doi.org/10.1007/s11661-018-4705-1.

[28] M.S. Titus, R.K. Rhein, P.B. Wells, P.C. Dodge, G.B. Viswanathan, M.J. Mills, A. Van der Ven, T.M. Pollock, Solute segregation and deviation from bulk thermodynamics at nanoscale crystalline defects, Sci. Adv. 2 (2016) e1601796. https://doi.org/10.1126/sciadv.1601796.

[29] M. Lenz, Y.M. Eggeler, J. Müller, C.H. Zenk, N. Volz, P. Wollgramm, G. Eggeler, S. Neumeier, M. Göken, E. Spiecker, Tension/Compression asymmetry of a creep deformed single crystal Co-base superalloy, Acta Mater. 166 (2019) 597–610. https://doi.org/10.1016/j.actamat.2018.12.053.

[30] T.M. Smith, B.D. Esser, B. Good, M.S. Hooshmand, G.B. Viswanathan, C.M.F. Rae, M. Ghazisaeidi, D.W. McComb, M.J. Mills, Segregation and Phase Transformations Along Superlattice Intrinsic Stacking Faults

in Ni-Based Superalloys, Metall. Mater. Trans. A. 49 (2018) 4186–4198. https://doi.org/10.1007/s11661-018-4701-5.

[31] D. Barba, T.M. Smith, J. Miao, M.J. Mills, R.C. Reed, Segregation-Assisted Plasticity in Ni-Based Superalloys, Metall. Mater. Trans. A. 49 (2018) 4173–4185. https://doi.org/10.1007/s11661-018-4567-6.

[32] D. Barba, E. Alabort, S. Pedrazzini, D.M. Collins, A.J. Wilkinson, P.A.J. Bagot, M.P. Moody, C. Atkinson, A. Jérusalem, R.C. Reed, On the microtwinning mechanism in a single crystal superalloy, Acta Mater. 135 (2017) 314–329. https://doi.org/10.1016/j.actamat.2017.05.072.

[33] V.A. Vorontsov, L. Kovarik, M.J. Mills, C.M.F. Rae, High-resolution electron microscopy of dislocation ribbons in a CMSX-4 superalloy single crystal, Acta Mater. 60 (2012) 4866–4878. https://doi.org/10.1016/j.actamat.2012.05.014.

[34] M.S. Titus, A. Mottura, G. Babu Viswanathan, A. Suzuki, M.J. Mills, T.M. Pollock, High resolution energy dispersive spectroscopy mapping of planar defects in L1₂-containing Co-base superalloys, Acta Mater. 89 (2015) 423–437. https://doi.org/10.1016/j.actamat.2015.01.050.

[35] S.K. Makineni, A. Kumar, M. Lenz, P. Kontis, T. Meiners, C. Zenk, S. Zaefferer, G. Eggeler, S. Neumeier, E. Spiecker, D. Raabe, B. Gault, On the diffusive phase transformation mechanism assisted by extended dislocations during creep of a single crystal CoNi-based superalloy, Acta Mater. 155 (2018) 362–371. https://doi.org/10.1016/j.actamat.2018.05.074.

[36] A. Mottura, A. Janotti, T.M. Pollock, A first-principles study of the effect of Ta on the superlattice intrinsic stacking fault energy of L1₂-Co₃(Al,W), Intermetallics. 28 (2012) 138–143. https://doi.org/10.1016/j.intermet.2012.04.020.

[37] A. Mottura, A. Janotti, T.M. Pollock, Alloying effects in the γ' phase of Co-based superalloys, in: Superalloys 2012 Twelfth Int. Symp., TMS, 2012: pp. 685–693.

[38] P.J.H. Denteneer, W. van Haeringen, Stacking-fault energies in semiconductors from first-principles calculations, J. Phys. C Solid State Phys. 20 (1987) L883–L887. https://doi.org/10.1088/0022-3719/20/32/001.

[39] M. Chen, C.-Y. Wang, First-principles investigation of the site preference and alloying effect of Mo, Ta and platinum group metals in γ'-Co₃(Al,W), Scr. Mater. 60 (2009) 659–662. https://doi.org/10.1016/j.scriptamat.2008.12.040.

[40] J. He, C.H. Zenk, X. Zhou, S. Neumeier, D. Raabe, B. Gault, S.K. Makineni, On the atomic solute diffusional mechanisms during compressive creep deformation of a Co-Al-W-Ta single crystal superalloy, Acta Mater. 184 (2020) 86–99. https://doi.org/10.1016/j.actamat.2019.11.035.

[41] S. Neumeier, H.U. Rehman, J. Neuner, C.H. Zenk, S. Michel, S. Schuwalow, J. Rogal, R. Drautz, M. Göken, Diffusion of solutes in fcc Cobalt investigated by diffusion couples and first principles kinetic Monte Carlo, Acta Mater. 106 (2016) 304–312. https://doi.org/10.1016/j.actamat.2016.01.028.

[42] C.L. Fu, R. Reed, A. Janotti, M. Kremar, On the Diffusion of Alloying Elements in the Nickel-Base Superalloys, in: Superalloys 2004 Tenth Int. Symp., TMS, 2004: pp. 867–876. https://doi.org/10.7449/2004/Superalloys_2004_867_876.

[43] M.S.A. Karunaratne, P. Carter, R.C. Reed, Interdiffusion in the face-centred cubic phase of the Ni–Re, Ni–Ta and Ni–W systems between 900 and 1300°C, Mater. Sci. Eng. A. 281 (2000) 229–233. https://doi.org/10.1016/S0921-5093(99)00705-4.

[44] M.S.A. Karunaratne, R.C. Reed, Interdiffusion of the platinum-group metals in nickel at elevated temperatures, Acta Mater. 51 (2003) 2905–2919. https://doi.org/10.1016/S1359-6454(03)00150-8.

[45] C.A. Schneider, W.S. Rasband, K.W. Eliceiri, NIH Image to ImageJ: 25 years of image analysis, Nat. Methods. 9 (2012) 671–675. https://doi.org/10.1038/nmeth.2089.

[46] V. Caccuri, J. Cormier, R. Desmorat, γ'-Rafting mechanisms under complex mechanical stress state in Ni-based single crystalline superalloys, Mater. Des. 131 (2017) 487–497. https://doi.org/10.1016/j.matdes.2017.06.018.

[47] S.K. Makineni, M. Lenz, P. Kontis, Z. Li, A. Kumar, P.J. Felfer, S. Neumeier, M. Herbig, E. Spiecker, D. Raabe, B. Gault, Correlative Microscopy—Novel Methods and Their Applications to Explore 3D Chemistry and Structure of Nanoscale Lattice Defects: A Case Study in Superalloys, JOM. 70 (2018) 1736–1743. https://doi.org/10.1007/s11837-018-2802-7.

[48] M. Herbig, P. Choi, D. Raabe, Combining structural and chemical information at the nanometer scale by correlative transmission electron microscopy and atom probe tomography, Ultramicroscopy. 153 (2015) 32–39. https://doi.org/10.1016/j.ultramic.2015.02.003.

[49] G. Kresse, J. Hafner, Ab initio molecular-dynamics simulation of the liquid-metal–amorphous-semiconductor transition in germanium, Phys. Rev. B. 49 (1994) 14251–14269. https://doi.org/10.1103/PhysRevB.49.14251.

[50] G. Kresse, J. Furthmüller, Efficiency of ab-initio total energy calculations for metals and semiconductors using a plane-wave basis set, Comput. Mater. Sci. 6 (1996) 15–50. https://doi.org/10.1016/0927-0256(96)00008-0.

[51] G. Kresse, J. Furthmüller, Efficient iterative schemes for *ab initio* total-energy calculations using a plane-wave basis set, Phys. Rev. B. 54 (1996) 11169–11186. https://doi.org/10.1103/PhysRevB.54.11169.

[52] J.P. Perdew, K. Burke, M. Ernzerhof, Generalized Gradient Approximation Made Simple, Phys. Rev. Lett. 77 (1996) 3865–3868. https://doi.org/10.1103/PhysRevLett.77.3865.

[53] P.E. Blöchl, Projector augmented-wave method, Phys. Rev. B. 50 (1994) 17953–17979. https://doi.org/10.1103/PhysRevB.50.17953.

[54] H.J. Monkhorst, J.D. Pack, Special points for Brillouin-zone integrations, Phys. Rev. B. 13 (1976) 5188–5192. https://doi.org/10.1103/PhysRevB.13.5188.

[55] J. Koßmann, T. Hammerschmidt, S. Maisel, S. Müller, R. Drautz, Solubility and ordering of Ti, Ta, Mo and W on the Al sublattice in L12-Co3Al, Intermetallics. 64 (2015) 44–50. https://doi.org/10.1016/j.intermet.2015.04.009.

[56] S. Kibey, J.B. Liu, D.D. Johnson, H. Sehitoglu, Generalized planar fault energies and twinning in Cu–Al alloys, Appl. Phys. Lett. 89 (2006) 191911. https://doi.org/10.1063/1.2387133.

[57] L.-Y. Tian, R. Lizárraga, H. Larsson, E. Holmström, L. Vitos, A first principles study of the stacking fault energies for fcc Co-based binary alloys, Acta Mater. 136 (2017) 215–223. https://doi.org/10.1016/j.actamat.2017.07.010.

[58] B. Yin, Z. Wu, W.A. Curtin, Comprehensive first-principles study of stable stacking faults in hcp metals, Acta Mater. 123 (2017) 223–234. https://doi.org/10.1016/j.actamat.2016.10.042.

[59] A. Bezold, N. Volz, F. Xue, C.H. Zenk, S. Neumeier, M.G. Ken, On the Precipitation-Strengthening Contribution of the Ta-Containing Co3(Al,W)-Phase to the Creep Properties of g/gp Cobalt-Base Superalloys, Metall. Mater. Trans. A. (n.d.) 8.

[60] P.C.J. Gallagher, The influence of alloying, temperature, and related effects on the stacking fault energy, Metall. Trans. 1 (1970) 2429–2461.

[61] K.V. Vamsi, S. Karthikeyan, Yield anomaly in L12 Co3AlxW1−x vis-à-vis Ni3Al, Scr. Mater. 130 (2017) 269–273. https://doi.org/10.1016/j.scriptamat.2016.11.039.

[62] T. Murakumo, Y. Koizumi, K. Kobayashi, H. Harada, Creep strength of Ni-base single-crystal superalloys on the γ/γ′ tie line, in: Superalloys 2004 Tenth Int. Symp., 2004: pp. 155–162.

[63] W. Schneider, H. Mughrabi, Investigation of the creep and rupture behaviour ot the single-crystal nickel-base superalloy CMSX-4 between 800°C and 1100°C, in: Proc. 5th Int. Conf. Creep Fract. Eng. Mater. Struct., 1993: pp. 209–220.

[64] C. Carry, J.L. Strudel, Apparent and effective creep parameters in single crystals of a nickel base superalloy—II. Secondary creep, Acta Metall. 26 (1978) 859–870. https://doi.org/10.1016/0001-6160(78)90035-4.

[65] T.M. Smith, B.D. Esser, N. Antolin, A. Carlsson, R.E.A. Williams, A. Wessman, T. Hanlon, H.L. Fraser, W. Windl, D.W. McComb, M.J. Mills, Phase transformation strengthening of high-temperature superalloys, Nat. Commun. 7 (2016). https://doi.org/10.1038/ncomms13434.